# ENHANCED USAGE OF KEYS OBTAINED BY PHYSICAL, UNCONDITIONALLY SECURE DISTRIBUTIONS

LASZLO B. KISH [(1)], CLAES-GÖRAN GRANQVIST [(2)]

[(1)] *Department of Electrical Engineering, Texas A&M University, College Station, TX 77843-3128, USA*

*laszlo.kish@ece.tamu.edu*

[(2)] *Department of Engineering Sciences, The Ångström Laboratory, Uppsala University, P.O. Box 534, SE-75121 Uppsala, Sweden*

Unconditionally secure physical key distribution schemes are very slow, and it is practically impossible to use a one-time-pad based cipher to guarantee unconditional security for the encryption of data because using the key bits more than once gives out statistical information, for example via the known-plain-text-attack or by utilizing known components of the protocol and language statistics. Here we outline a protocol that reduces this speed problem and allows almost-one-time-pad based communication with an unconditionally secure physical key of finite length. The physical, unconditionally secure key is not used for data encryption but is employed in order to generate and share a new software-based key without any known-plain-text component. The software-only-based key distribution is then changed from computationally secure to unconditionally secure, because the communicated key-exchange data (algorithm parameters, one-way functions of random numbers, etc.) are encrypted in an unconditionally secure way with a one-time-pad. For practical applications, this combined physical/software key distribution based communication looks favorable compared to the software-only and physical-only key distribution based communication whenever the speed of the physical key distribution is much lower than that of the software-based key distribution. A mathematical security proof of this new scheme remains an open problem.

*Keywords:* information theoretic security, secure key, one-time-pad.

*Unconditionally secure physical key distribution* of a cryptographic system means that the eavesdropper's (Eve's) probability $p_E$ of successfully guessing the bits arbitrarily can approach the case of *perfect security*—i.e., $p_E = 0.5$—provided that sufficient resources, such as time, are available. It is important to note that this definition does not prescribe that the amount of required resources is practical and/or economical, but obviously the laws of physics must not be violated.

*Enhanced usage of keys by physical, unconditionally secure distributions*

Such a key distribution requires a specific physical system: either a quantum system providing a quantum key distribution [1] or a classical physical system such as the Kirchhoff-law–Johnson-noise (KLJN) protocol [2–5]. A common characteristic of these systems is that the speed of key exchange is low, and therefore it is practically impossible to utilize a *one-time-pad* based cipher, which is using any key bits only for a single occasion, to guarantee unconditional security for the encryption of data. Using the key bits more than once gives out statistical information, *e.g.* via the known-plain-text-attack [1], by employing known data components of the protocol or language statistics. Here we outline a protocol that reduces this problem and allows the use of one-time-pad based communication with unconditionally secure physical key distribution. In the rest of the paper, we refer to the unconditionally secure physical key as a *hardware based key* (HBK).

Our goal is to generate much longer keys than those provided by the HBK and to make it difficult to identify the HBK even when if it is used for encryption over an extended time. The new core protocol is devised as follows:

(*i*) First we share an HBK with an unconditionally secure physical key exchange protocol. If the HBK is not clean enough and the key exchange has low error probability, we can make the HBK sufficiently clean by privacy amplification, for example by use of the simplest XOR type [6] which does not introduce correlations among bits. For example, KLJN has astronomically low error probability [5] and therefore is very favorable for privacy amplification [6]. The resulting HBK of length $N$ can arbitrarily approach perfect security provided enough time and/or other resources are available [2,6]; thus it is indeed unconditionally secure.

(*ii*) Then we use the Diffie–Hellman–Merkle (DHM) method [7], or some other *computationally secure* (*e.g.*, a one-way-function based) technique, to generate a *software based key* (SBK) with length $M >> N$. This is accomplished so that the computationally secure method is employed via unconditionally secure communication, utilizing the HBK as a one-time-pad to generate a $K$-long key where, obviously, $K < N$. Then we repeat the process independently $M/K$ times to get an $M$-long SBK.

(*iii*) For example, the DHM method normally uses a publicly agreed joint prime number, spontaneously generated random numbers, and functions of these, that are publicly communicated. In our method, all of these key exchange steps are encrypted by the HBK based one-time-pad, including the selection of the joint prime number, which in this way is randomly selected and not public. If the communicated numbers during this process were independent, then the resulting SBK key would be perfectly secure. However, there are algorithmic relations between some of these numbers (note, the one-way function values are hard to invert even if publicly known) and thus, theoretically, there can be some information leak in the process. However it is extremely difficult for Eve to extract any useful information because, due to the one-time-pad encryption, she does not know any of the numbers of the protocol, not even those that are public knowledge in the regular DHM method. Thus neither the input nor the output of the one-way functions would be known by Eve, and therefore any information about them would have to come in some indirect way while the protocol is repeated by new random numbers. It then follows that—even if Eve had a hypothetical, efficient special-purpose algorithm or a proper quantum computer or a noise-based logic engine, and would be able crack the classical DHM protocol—she would not have the publicly communicated numbers necessary to start with because all of them are encrypted.

(*iv*) The SBK is used as a one-time-pad for data communication. The potentially great length of the obtained SBK can allow its one-time-pad applications to encrypt messages.

The combined physical/software key distribution based communication, described above, looks favorable with regard to practical applications compared to the *software-only* and *physical-only* key distribution based communication methods whenever the speed of the physical key distribution is lower than that of the software-based key distribution.

A security proof for this novel scheme is an open problem. For example, what is the $M$ ($\gg N$) value at which brute force attempts to crack the original HBK become less favorable than possible algorithmic methods utilizing possible relations between the encrypted numbers mentioned above? Further work is needed to give an answer to this question.

## References

1. Horace P. Yuen, "Simple explanation on why QKD keys have not been proved secure"; http://arxiv.org/abs/1408.4780 (2014).
2. L.B. Kish, C.G. Granqvist, "On the security of the Kirchhoff-law–Johnson-noise (KLJN) communicator", *Quantum Information Processing* **13** (2014) 2213–219. DOI: 10.1007/s11128-014-0729-7; http://arxiv.org/abs/1309.4112; http://vixra.org/abs/1309.0106.
3. L.B. Kish, "Totally secure classical communication utilizing Johnson(-like) noise and Kirchoff's law", *Physics Letters A* **352** (2006) 178–182.
4. R. Mingesz, Z. Gingl, L.B. Kish, "Johnson(-like)-noise–Kirchhoff-loop based secure classical communicator characteristics, for ranges of two to two thousand kilometers, via model-line", *Physics Letters A* **372** (2008) 978–984.
5. Y. Saez, L.B. Kish, "Errors and their mitigation at the Kirchhoff-law–Johnson-noise secure key exchange", *PLoS ONE* **8** (2013) e81103.
6. T. Horvath, L.B. Kish, J. Scheuer, "Effective privacy amplification for secure classical communications", *EPL (Europhysics Letters)* **94** (2011) 28002-p1–p6; http://arxiv.org/abs/1101.4264.
7. M.E. Hellman, B.W. Diffie, R.C. Merkle, "Cryptographic apparatus and method", US Patent 4,200,770, 29 April 1980; http://www.google.com/patents/US4200770.